\documentclass{ws-ijmpe}
\usepackage[super,compress]{cite}
\usepackage{graphicx}
\usepackage{float}
\restylefloat{table}
\usepackage{slashed}
\usepackage{mathtools}
\usepackage{dcolumn}
\usepackage{longtable}
\RequirePackage{graphicx}
\RequirePackage{mathptmx}      
\RequirePackage{flushend}
\RequirePackage[numbers,sort&compress]{natbib}
\RequirePackage[colorlinks,citecolor=blue,urlcolor=blue,linkcolor=blue]{hyperref}
\DeclarePairedDelimiter\bra{\langle}{\rvert}
\DeclarePairedDelimiter\ket{\lvert}{\rangle}
\DeclarePairedDelimiterX\braket[2]{\langle}{\rangle}{#1 \delimsize\vert #2}
\DeclarePairedDelimiterX\inner[2]{\langle}{\rangle}{#1,#2}

\begin{document}


\catchline{}{}{}{}{}

\title{Effects of coupled channels on $c\bar{b}$ masses and decays in NRQM with OGEP
}
\author{Manjunath Bhat}

\address{Department of Physics, St Philomena college, 
              Darbe, Puttur  574 202, India\\
manjunathbhat61@yahoo.in}
\author{Antony Prakash Monteiro
}

\address{ Department of Physics, St Philomena college, 
              Darbe, Puttur  574 202, India\\
aprakashmonteiro@gmail.com}

\author{ K. B. Vijaya Kumar}

\address{Department of Physics, Mangalore University,
Mangalagangothri P.O., Mangalore - 574199, India\\
kbvijayakumar@yahoo.com}
\maketitle

\begin{history}
\received{Day Month Year}
\revised{Day Month Year}
\end{history}

\begin{abstract}
The complete spectrum of $c\bar{b}$ states is obtained in a phenomenological non relativistic quark model(NRQM), which consists of a confinement potential and one gluon exchange potential (OGEP) as effective quark - antiquark potential with coupled channel effects. We make predictions for the radiative decay (E1 and M1) widths and weak decay widths of $c\bar{b}$ states in the framework of NRQM formalism.

\keywords{Mesons; Phenomenological quark models; Non relativistic quark models; Leptonic; semileptonic; radiative decays of mesons, coupled channel effects}
\end{abstract}

\ccode{PACS: 14.40.-n;12.3.-x,12.39.-Jh,13.20.-v}


\section{INTRODUCTION}

\label{sec:intro}

The investigation of masses of $c\bar{b}$ states gives us an opportunity to obtain information on the nature of the strong interaction thereby it throws up an interesting issue and a tantalizing problem. Since the charmed bottom meson $c\bar{b}$ is an intermediate state of the $c\bar{c}$ and $b\bar{b}$ mesons, its analysis could give detailed information on the balance between perturbative and non perturbative effects. There are a good number of theoretical models that study leptonic, semi leptonic and hadronic decay channels of $c\bar{b}$ states. Using NRQM formalism we have already studied mass spectra and decay properties of $c\bar{b}$ meson. This work attempts to study the effects of coupled channels on $c\bar{b}$ masses and its decays in NRQM.\\

The NRQM formalism is found to provide systematic treatment of the perturbative and non perturbative components of QCD at hadronic scale. The masses of the $c\bar{b}$ states are predicted using NRQM whose parameters are tuned to produce the spectra of the observed charmonium and bottomonium states.\\

The paper is organized in 4 sections. In sec.~\ref{sec:TB} we give the description of our model in the theoretical background, the framework of the coupled-channel analysis and  description of various decays. In sec.~\ref{sec:RD} we discuss the results and the conclusions are drawn in sec.~\ref{sec:C} with a comparison to other models. \\ 

\section{THEORETICAL BACKGROUND}
\label{sec:TB}
\subsection{The Hamiltonian}
In our model we use the Hamiltonian which includes includes kinetic energy, confinement potential and one gluon potential (OGEP)\cite {VH04,BS08, VB09,DG75}.
\begin{equation}
H=K+V_{CONF}+V_{OGEP}
\end{equation}
 where $K$ is the kinetic energy part, $V_{CONF}$ is confinement potential that comes from the non perturbative treatment of QCD, $V_{OGEP}$ is the residual interaction from perturbative treatment of quark-antiquark system.
\subsection{The Linear Confinement Potential}
In literature different types of confinement potentials are chosen depending upon the unique features of the phenomenological quark model under consideration. They can be harmonic oscillator potential ($V\sim r^2$) or logarithmic potential ($V\sim ln(r)$) or linear potential ($V\sim r$). We deem it fit to choose linear potential that represents non perturbative effect of QCD that explains quark confinement within the color singlet system ~\cite{BS08, VB09}.

\begin{equation}
V_{CONF}(\vec{r}_{ij})= -a_{c}r_{ij}\vec{\lambda}_{i}\cdot
\vec{\lambda}_{j} \label{eq:V-conf}
\end{equation}
where $a_{c}$ is the confinement strength, $\lambda_{i}$ and $\lambda_{j}$ are
the generators of the color SU(3) group for the $ i^{\rm th}$ and $ j^{\rm th}$ quarks. Since confinement is of two body system we leave out the spin-orbit contribution for it adds nothing practically to the interaction.\\
\subsection{The Short Distance Behaviour}
The one gluon exchange potential(OGEP) describes the short distance behavior. 
The central part of the two-body potential due to OGEP is ~\cite{DG75},
\begin{equation}
V_{OGEP}(\vec{r}_{ij})=\frac{\alpha_s}{4}\vec{\lambda}_i\cdot\vec{\lambda}_j\left[\frac{1}{r_{ij}}-\frac{\pi}{M_iM_j}\left(\frac{M_i}{M_j}+\frac{M_j}{M_i}+\frac{2}{3}\vec{\sigma}_i\cdot\vec{\sigma}_j\right)\delta(\vec{r}_{ij})\right]
\label{eq:V-ogep}
\end{equation}
where the first term represents the residual Coulomb energy and the
second term is the chromo-magnetic interaction leading to the
hyperfine splitting. $\sigma_{i}$ is the Pauli spin operator and
$\alpha_{s}$ is the quark-gluon coupling constant.

\hspace{-0.5cm}The spin-orbit interaction of OGEP is given by,
\begin{equation}
V^{SO}_{OGEP}(\vec{r})=-\frac{\alpha_s}{4}{\bf \lambda_i\cdot\lambda_j}\left[\frac{3}{8M_iM_j}\frac{1}{r^3}(\vec{r}\times\vec{p})\cdot({\bf \sigma_i}+{\bf \sigma}_j)\right]
\end{equation}
The following tensor term \cite{BA89, VB95} is considered, 
\begin{equation}
V^{ten}_{OGEP}(\vec{r})=-\frac{\alpha_s}{4}{\bf \lambda_i\cdot\lambda_j}\left[\frac{1}{4M_iM_j}\frac{1}{r^3}\right]\hat{S}_{ij}
\end{equation}
where,
\begin{equation}
\hat{S}_{ij}=[3(\vec{\sigma}_i\cdot\hat{r})(\vec{\sigma}_j\cdot\hat{r})-\vec{\sigma}_i\cdot\vec{\sigma}_j]
\end{equation}

\subsection{Coupled Channel Effects}
 
The coupled channel effects (hadronic loop effects) have been neglected by most of the QCD inspired potential models in calculating the masses of mesons. The $BD$, $B_sD_s$ etc., channels strongly couple to the $c\bar{b}$ states. These channels give rise to mass shifts both below and above BD meson pair creation threshold. Also above threshold  these effects lead to the strong decay of $B_c$ meson.  These effects in our calculation are introduced explicitly  through the $^3P_0$ pair creation model for the decay of meson $A\to B+C$ which was proposed by Micu and developed by Le Yaouanc {\textit{et al}} and others\cite{MI69,Li12,TS08,ES96}. The main assumption of the model is that the strong decay of meson $A$ takes place through the creation of a pair of quark and anti-quark from vacuum with quantum number $J^{PC}=0^{++}$. The created quark anti-qurak pair recombines with the quark and anti-qurak in the initial meson state forming final meson states i.e, mesons B and C. \cite{TO79,ON84,TO95,TO96,BE83,KH84,BE80,MI69,Li12,TB97,TS08,ES96,TO85,PG91,HZ91}.\\

\hspace{-0.5cm}In the coupled channel model, the full hadronic state is given by \cite{Li12,TS08,ES96}
\begin{equation}
\ket{\psi}=\ket{A}+\sum_{BC}\ket{BC}
\end{equation}
for open flavour strong decay $A\to BC$. Here A, B, C denote mesons. 

\hspace{-0.5cm}The wave function $\ket{\psi}$ obeys the equation
\begin{equation}
 H\ket{\psi}=M\ket{\psi}
\end{equation}
The Hamiltonian H for this combined system consists of a valence Hamiltonian $H_0$ and an interaction Hamiltonian $H_I$ which couples the valence and continuum sectors.
The matrix element of the valence-continuum coupling Hamiltonian is given by \cite{TS08,ES96} 
\begin{equation}
\bra{BC}H_I\ket{A}=h_{fi}\delta (\vec{P}_A-\vec{P}_B-\vec{P}_C)
\end{equation}
where $h_{fi}$ is the decay amplitude.

\hspace{-0.5cm}The mass shift of meson A due to its continuum coupling to BC can be expressed in terms of partial wave amplitude ${\cal M}_{LS}$ \cite{Li12,ES96}
\begin{eqnarray*}
\Delta M_A^{(BC)}=\int^\infty_0dp\frac{p^2}{E_B+E_C-M_A-i\epsilon}\int d\Omega_p|h_{fi}(p)|^2\\
\quad=\int^\infty_0dp\frac{p^2}{E_B+E_C-M_A-i\epsilon}\sum_{LS}|{\cal M}_{LS}|^2
\end{eqnarray*}
\begin{equation}
\Delta M_A^{(BC)}={\cal P}\int^\infty_0dp\frac{p^2}{E_B+E_C-M_A}\sum_{LS}|{\cal M}_{LS}|^2+i\pi\left(\frac{p*E_B*E_C}{M_A}\sum_{LS}|{\cal M}_{LS}|^2\right)|_{E_B+E_C=M_A}
\end{equation}
The decay amplitude $h_{fi}$ can be combined with relativistic phase space to give the differential decay rate, which is
\begin{equation}
\frac{d\Gamma_{A\to BC}}{d\Omega}=2\pi P\frac{E_BE_C}{M_A}|h_{fi}|^2
\end{equation}
where in the rest frame of A, we have $\vec{P}_A=0$ and $P=|\vec{P}_B|=|\vec{P}_C|$,
and 
\begin{equation}
P=\sqrt{[M^2_A-(M_B+M_C)^2][M^2_A-(M_B-M_C)^2]}/(2M_A)
\end{equation}
Finally the total decay rate is given by \cite{Li12,ES96}
\begin{equation}
\label{strong}
\Gamma_{A\to BC}=2\pi P\frac{E_BE_C}{M_A}\sum_{LS}|{\cal M}_{LS}|^2
\end{equation}

\subsection{Decay Properties}  

\subsubsection{Electric Dipole (E1) Transitions} 

The partial width for electric dipole (E1) transitions is given by
\begin{equation}
\Gamma_{(i\to f+\gamma)}=(2J'+1)\frac{4}{3}Q^2_b\alpha k^3_0S_{if}^E\left|{\cal E}_{if}\right|^2 
\end{equation}
Here $k_0$ is the energy of the emitted photon and it is given by
$k_0=\frac{m^2_a-m^2_b}{2m_a}$. 
$\alpha$ is the fine structure constant. $Q_b=1/3$ is the charge of the b quark in units of $|e|$,  the statistical factor $S_{if}^E=\rm{max}(l,l')\left \{\begin{array}{ccc}
J & 1 & J' \\
l' & s  & l 
 \end{array} \right\}^2$, 
$J,~J'$ are the total angular momentum of initial and final mesons, $l,~l'$ are the orbital angular momentum of initial and final mesons and $s$ is the spin of initial meson.
 \begin{equation}
 {\cal E}_{if}=\frac{3}{k_0}\int^\infty_0 r^3R_{nl}(r)R'_{nl}(r)dr\left[\frac{k_0r}{2}j_0\left(\frac{k_0r}{2}\right)-j_1\left(\frac{k_0r}{2}\right)\right]
 \end{equation}
is the radial overlap integral which has the dimension of length, with $R_{nl}(r)$ being the normalized radial wave functions for the corresponding states.
 
\subsubsection{Magnetic Dipole (M1) Transitions}
The partial decay width for M1 transitions is \cite{WJ88,NL78,CE05,WE86,BB95,NB99,BA96}
\begin{equation}
\label{decay}
\begin{split}
\Gamma_{a\to b+\gamma}=\delta_{L_aL_b}4\alpha k^3_0\frac{E_b(k_0)}{m_a}\left(\frac{Q_c}{m_c}+(-1)^{S_a+S_b}\frac{Q_{\bar{b}}}{m_{\bar{b}}}\right)^2(2S_a+1)
\times(2S_b+1)(2J_b+1)\\
\left \{\begin{array}{ccc}
S_a & L_a & J_a \\
J_b & 1  & S_b 
 \end{array} \right\}^2\left \{\begin{array}{ccc}
1 & \frac{1}{2} & \frac{1}{2} \\
\frac{1}{2} & S_a  & S_b 
 \end{array} \right\}^2
\times\left[\int^\infty_0  R_{n_bL_b}(r)r^2j_0(kr/2)R_{n_aL_a}(r) dr\right]^2
\end{split}
\end{equation} 
where $\int^\infty_0 dr  R_{n_bL_b}(r)r^2j_0(kr/2)R_{n_aL_a}(r)$ is the overlap integral for unit operator between the coordinate wave functions of the initial and the final meson states, $j_0(kr/2)$ is the spherical Bessel function, $m_b$ is the mass of bottom quark. $J_b$ is the total angular momentum of final meson state. \\

\subsubsection{Weak Decays}
\label{weak}
 Weak decays of $B_c$ meson plays a special role in our understanding of physics of the Standard Model and beyond. Various diagrams can contribute to the weak decays at the quark level. These are
mainly a) Spectator quark, b) W-exchange, c) W-annihilation and d)
Penguin diagrams. Due to the helicity and color considerations, the W-exchange diagrams are suppressed. Penguin diagrams are also expected to be small in strength. Hence the dominant quark level processes seem to be the process in which one of the quarks(anti-quark) behave like spectator and the W-annihilation\cite{ALT82,ALI91}. Using this picture, after evaluating the contributing diagrams we get the decay widths for a hadron containing a b quark or c quark as in eqns(18, 19 and 20).\\

 A rough estimate of the $B_c$ weak decay widths can be done by treating the $\bar{b}$-quark and $c$-quark decays independently so that $B_c$ decays can be divided into three classes \cite{AA99,GS91}$\colon$ (i)the $\bar{b}$-quark decay with spectator $c$-quark, (ii) the $c$-quark decay with spectator $\bar{b}$-quark, and (iii) the annihilation $B^+_c\rightarrow l^+\nu_l$ ($c\bar{s},~u\bar{s}$), where $l=e,~\mu,~\tau$.  
The total decay width can be written as the sum over partial widths
 \begin{equation}
 \Gamma(B_c\rightarrow X)=\Gamma_1(\bar{b}\rightarrow X)+\Gamma_2(c\rightarrow X)+\Gamma_3(ann)
 \end{equation}
 In the spectator approximation:
 \begin{equation}
 \Gamma_1(\bar{b}\rightarrow X)=\frac{9G^2_F|V_{cb}|^2m^5_b}{192\pi^3}\label{eq1}
 \end{equation}
  and
\begin{equation}
 \Gamma_2(c\rightarrow X)=\frac{5G^2_F|V_{cs}|^2m^5_c}{192\pi^3} \label{eq2}
 \end{equation}
 In the above expressions $V_{cb}$ and $V_{cs}$ are the elements of the CKM matrix, $G_F=1.16637\times 10^{-5}$ is the Fermi coupling constant, $m_c$ and $m_b$ are the masses of c and b quarks respectively. \\
 
The decay of vector meson into charged leptons proceeds through the virtual photon $(q\bar{q}\rightarrow l^+l^-)$. The $^3S_1$ and $^3D_1$ states have quantum numbers of a virtual photon, $J^{PC}=1^{--}$ and can annihilate into lepton pairs through one photon.
Annihilation widths such as $c\bar{b}\rightarrow l\nu_l$ are given by the expression
\begin{equation}
\Gamma_3(ann)=\frac{G^2_F}{8\pi}|V_{bc}|^2f^2_{B_c}M_{B_c}\sum_i m^2_i\left(1-\frac{m^2_i}{M^2_{B_c}}\right)C_i\label{eq3}
\end{equation}
where $m_i$ is the mass of the heavier fermion in the given decay channel. For lepton channels $C_i=1$ while for quark channels $C_i=3|V_{q\bar{q}}|^2$. \\
The pseudo scalar decay constant $f_{B_c}$ is defined by \cite{EC94}
\begin{equation}
\bra{0}\bar{b}(x)\gamma^\mu\gamma_5 c(x)\ket{B_c(k)}=if_{B_c}V_{cb}k^\mu 
\end{equation}  
 where $k^\mu$ is the four-momentum of the $B_c$ meson.
 In the non relativistic limit the pseudo scalar decay constant is proportional to the wave function at the origin and is given by van Royen-Weisskopf formula \cite{RV67}
 \begin{equation}
 f_{B_c}=\sqrt{\frac{12}{M_{B_c}}}\psi(0)
 \end{equation}
 Here $\psi(0)$ is wavefunction at the origin.

\section{Results and Discussion}
\label{sec:RD}
\subsection{Mass Spectra}
The parameters used in our model are listed in Table \ref{para}. We have fixed the parameters using the approach used in our earlier works\cite{VH04,VB09,KB97}.  We obtain the parameter $b$ by minimizing the expectation value of the Hamiltonian i.e, $\frac{\partial \bra{\psi}H\ket{\psi}}{\partial b}=0$. The confinement strength $a_c$ is fixed by the stability condition for variation of mass of the vector meson($B_c^*$ meson) against the size parameter $b$.  We initially assume a set of values for the parameters  $\alpha_s$, $m_b$, $m_c$ and we consturct a $5\times 5$ matrix and diagonalize the matrix to obtain mass of $B_c$ meson states. Then we tune these parameters to obtain an agreement with the experimental value for the mass of $B_c$ meson. 
\begin{table}[!h]
\centering
\tbl{\label{para}\bf Parameters of the model}
{\begin{tabular*}{\linewidth}{@{\extracolsep{\fill}}lrrrrrrrrl@{}}
\hline
$m_c$~(MeV)&1525.0\\
$m_b$~(MeV)&4825.0\\
$b$~(fm)&0.350\\
$\alpha_s$& 0.3\\
$a_c~{\rm MeV~fm^{-1}}$&175\\
\hline
\end{tabular*}}
\end{table}

\begin{table*}[!h]
\centering
\caption{\label{Shift}\bf Mass shifts (in MeV).}
\begin{tabular*}{\textwidth}{@{\extracolsep{\fill}}lrrrrrrrrl@{}}
\hline
Bare $c\bar{b}$ State &&\\
$n~^{2S+1}L_J$&\multicolumn{1}{c}{BD}&\multicolumn{1}{c}{$B_sD_s$}&\multicolumn{1}{c}{$B_0D_0$}&\multicolumn{1}{c}{$B^*D$}&\multicolumn{1}{c}{$B^*_sD_s$}&\multicolumn{1}{c}{$B^*D^*$}&\multicolumn{1}{c}{$B^*_sD^*_s$}&\multicolumn{1}{c}{Total}\\
\hline
$1~^1S_0$&0&0&0&-5.661&-5.033&-10.434&-9.328&-30.456\\

$1~^3S_1$&-2.046&-1.805&-2.052&-3.955&-3.496&-7.293&-6.488&-27.135\\
$1~^3P_0$&-57.922&-57.406&-57.946&0&0&-19.088&-18.932&-211.294\\
$1~^1P_1$&0&0&0&-18.49&-18.393&-37.603&-37.901&-112.387\\
$1~^3P_1$&0&0&0&-38.390&-38.049&0&0&-76.439\\
$1~^3P_2$&-40.618&-40.314&-40.632&0&0&0&0&-121.557\\
$2~^1S_0$&0&0&0&-1.547&-1.361&-2.837&-2.523&-8.268\\
$2~^3S_1$&-0.546&-0.476&-0.548&-1.929&-1.711&-1.049&-0.920&-7.179\\
$1~^3D_1$&-30.675&-30.312&-30.682&-15.326&-15.146&-3.077&-3.044&-128.262\\
$1~^1D_2$&0&0&0&-3.147&-3.111&-27.643&-27.49&-61.391\\
$1~^3D_2$&0&0&0&-27.214&-27.552&-69.486&-68.957&-193.209\\
$1~^3D_3$&-40.753&-40.359&-40.772&-54.308&-53.783&-20.835&-20.606&-230.663\\
$2~^3P_0$&-148.72&-146.395&-148.828&0&0&-48.589&-47.903&-540.435\\
$2~^1P_1$&0&0&0&-25.081&-24.744&-49.343&-48.741&-147.909\\
$2~^3P_1$&0&0&0&-98.623&-97.088&0&0&-195.711\\
$2~^3P_2$&-79.114&-77.890&-79.171&0&0&0&0&-236.175\\

\hline
\end{tabular*}
\end{table*} 

\begin{table}[!h]
\caption{\label{spectrum}\bf $B_c$ meson mass spectrum (in MeV).}
\begin{tabular*}{\linewidth}{@{\extracolsep{\fill}}lccccccccccccl@{}}
\hline
State &&\\
$n~^{2S+1}L_J$&\multicolumn{1}{c}{This work}&\multicolumn{1}{c}{Ref.\cite{SJ96}}&\multicolumn{1}{c}{Ref. \cite{VA95}}&\multicolumn{1}{c}{Ref. \cite{ZV95}}&\multicolumn{1}{c}{Ref. \cite{EC94}}&\multicolumn{1}{c}{Ref.\cite{DR03}}&\multicolumn{1}{c}{Ref.\cite{SN85}}&\multicolumn{1}{c}{Ref.\cite{FL}}\\
\hline
$1~^1S_0$&6276&6247&6253&6260&6264&6270&6271&6286\\

$1~^3S_1$&6347&6308&6317&6340&6337&6332&6338&6341\\
$1~^3P_0$&6654&6689&6683&6680&6700&6699&6706&6701\\
$1P$&6683&6738&6717&6730&6730&6734&6741&6737\\
$1P'$&6729&6757&6729&6740&6736&6749&6750&6.760\\
$1~^3P_2$&6732&6773&6743&6760&6747&6762&6768&6772\\
$2~^1S_0$&6853&6853&6867&6850&6856&6835&6855&6882\\
$2~^3S_1$&6881&6886&6902&6900&6899&6881&6887&6914\\
$1~^3D_1$&6990&&7008&7010&7012&7072&7028&7019\\
$1D$&6985&&7001&7020&7012&7077&7041&7028\\
$1D'$&7010&&7016&7030&7009&7079&7036&7028\\
$1~^3D_3$&7021&&7007&7040&7005&7081&7045&7032\\
$2~^3P_0$&7107&&7088&7100&7108&7091&7122&\\
$2P$&7123&&7113&7140&7135&7126&7145&\\
$2P'$&7128&&7124&7150&7142&7145&7150&\\
$2~^3P_2$&7136&&7134&7160&7153&7156&7164&\\
\hline
\end{tabular*}
\end{table}
We evaluate the bare state masses and shifts due to $BD$, $B_sD_s$, $B^0D^0$, $B^*D$, $B^*_sD_s$, $B^*D^*$ and $B^*_sD^*_s$ loops (with $M_B=5279.26 ~\rm{MeV}$, $M_{B_s}=5366.77~ \rm{MeV}$, $M_{B^0}=5279.58~ \rm{MeV}$, $M_{B^*}=5324.6 ~\rm{MeV}$, $M_{B^*_s}=5415.4~ \rm{MeV}$, $M_{D}=1869.61~ \rm{MeV}$, $M_{D_s}=1968.30~ \rm{MeV}$, $M_{D_0}=1864.84~\rm{MeV}$, $M_{D^*}=2006.96~\rm{MeV}$ and $M_{D^*_s}=2112.1$ MeV).\\
  
   We consider the mixing between $^3P_1$ and $ ^1P_1$ and also between  $^3D_2$ and $^1D_2$ eigenstates due to the spin-orbit interaction terms. The mixing yields the $B_c$ mesons  with $J=1$ and $J=2$ states $P1$, $P1'$, $D2$ and $D2'$. These states are in general represented as
 \begin{eqnarray}
&\ket{nL'}=\ket{n~^1L_J}\cos\theta_{nL}+\ket{n~^3L_J}\sin\theta_{nL}\\
&\ket{nL}=-\ket{n~^1L_J}\sin\theta_{nL}+\ket{n~^3L_J}\cos\theta_{nL}
\end{eqnarray}
$$~J=L=1,2,3,\cdots$$
where $\theta_{nL}$ is a mixing angle, and the primed state has the heavier mass. For $L=J=1$ we have mixing of P states, with mixing angles $\theta_{1P}=0.4^\circ$ and $\theta_{2P}=0.05^\circ$.
Similarly for $L=J=2$ we have mixing of D states, with mixing angle 
$\theta_{1D}=0.05^\circ$. \\

 Table ~\ref{spectrum} shows the results for the masses of the $c\bar{b}$ states. The calculated masses are compared with other theoretical models and with available experimental data. Overall we obtain a good fit to the spectrum. The hyperfine mass splitting of singlet and triplet states $m(n~^3S_1)-m({}^1S_0)$ can shed light on the spin dependence of the energy levels. We obtain a hyperfine splitting of 71 MeV which is in good agreement with the other theoretical models. This difference is justified by calculating the $^3S_1-{}^1S_0$ splitting of the ground state which is given by 
 \begin{equation}
 M({}^3S_1)-M({}^1S_0)=\frac{32\pi\alpha_s|\psi(0)|^2}{9m_cm_b}
 \end{equation}
 We predict a mass of 6853 MeV for the first radial excitation $B_c$(2S) which is in good agreement with the experimental value  6842$\pm$4$\pm$5 MeV of $B_c$(2S)\cite{AG14}. The first radial excitation $B_c$(2S) is heavier than $B_c$(1S) by 577 MeV. The hyperfine splitting of $2S$ states is 28 MeV. The difference between the $B^*_c$(2S) and $B^*_c$(1S) masses turns out to be 534 MeV.

\subsection{Decay Properties}
The dominant multipole transitions $E1$ and $M1$ have been  studied and this helps us to extract information about new meson states and discover them. Radiative transitions are very important and interesting because the charge structure of the mesons and their quantum numbers can be determined  through these transitions. We consider $E1$ and $M1$ radiative transitions non relativistically for $B_c$ meson states. This potential model approach provides deatiled predictions which are further compared with experimental data. The possible $E_1$ decay modes listed in Table \ref{E1} are calculated and values of widths are given in the same. Though most of the predictions qualitatively agree with other theoretical models, some differ. These differences are due to different phase spaces arising from the different mass predictions. Wavefunction effects also play a major role in determining decay widths. The choice of $^3P_1-^1P_1$ mixing angles in different models is also a cause for the significant difference between the theoretical models in case of transitions involving $P1$ and $P1'$ states.  \\

 The radiative M1 transition rates of $B_c$ meson states are calculated and the results are presented in Table \ref{M1}. The M1 decay widths for allowed transitions (n$^3S_1 \rightarrow n' {^1S_0} + \gamma$,
$n = n'$ ) have been calculated and are compared with other non relativistic quark models \cite{EC94,SA95,FL}.  The decay widths of  hindered transitions($n\neq n'$) are zero in the non relativistic limit due to the orthogonality of the initial and final state wave functions. The hindered $M1$ transition rates are enhanced in this model by incorporating relativistic effects to the wavefuncion. \\

We have calculated weak decay widths of $B_c$ meson. The decay widths are calculated using $|V_{bc}|=0.044$ \cite{PDG} and $|V_{cs}|=0.975$ \cite{PDG}.  Calculated values of $\Gamma_1(\bar{b}\rightarrow X)$ is $9.628\times 10^{-4}~\rm{eV}$, $\Gamma_2(c\rightarrow X)$ is $7.712\times 10^{-4}~\rm{eV}$ and $\Gamma_3$ is $3.56\times 10^{-6}~\rm{eV}$.
Adding these results we get the total decay width~~ $\Gamma(\rm{total})=\Gamma_1+\Gamma_2+\Gamma_3=18.104\times 10^{-4}~\rm{eV}$ corresponding to a life time of $\tau=0.364~\rm{ps}$. The values of decay constant in various theoretical models are listed in Table \ref{fbc} and in Table \ref{T1} we compare the life time of $B_c$ meson calculated in our model with other models.\\
 \begin{table*}[!h]
\caption{\label{E1}\bf E1 transition rates of $B_c$ meson.}
{\begin{tabular*}{\textwidth}{@{\extracolsep{\fill}}lrrrrrrrrrrl@{}}
\hline
Transition&\multicolumn{1}{c}{k$_0$}&\multicolumn{1}{c}{This Work}&\multicolumn{1}{c}{Ref. \cite{DR03}}&\multicolumn{1}{c}{Ref. \cite{EC94}}&\multicolumn{1}{c}{Ref. \cite{VA95}}&\multicolumn{1}{c}{Ref.\cite{FL}}\\
&MeV&keV&keV&keV&keV&keV\\
\hline
$1^3P_0\rightarrow 1^3S_1\gamma$&307&30.7&75.5&79.2&65.3&74.2\\
$1P\rightarrow 1^3S_1\gamma$&336&49.4&87.1&99.5&77.8&75.8\\
$1P'\rightarrow 1^3S_1\gamma$&382&74.3&13.7&0.1&8.1&26.2\\
$1^3P_2\rightarrow 1^3S_1\gamma$&385&112.7&122&112.6&102.9&126\\
$1P\rightarrow 1^1S_0\gamma$&407&31.9&18.4&0&11.6&32.5\\
$1P'\rightarrow 1^1S_0\gamma$&453&44.5&147&56.4&131.1&128\\
$2^3S_1\rightarrow 1^3P_0\gamma$&227&8.0&5.53&7.8&7.7&9.6\\
$2^3S_1\rightarrow 1P\gamma$&198&8.5&7.65&14.5&12.8&13.3\\
$2^3S_1\rightarrow 1P'\gamma$&152&3.9&0.74&0&1.0&2.5\\
$2^3S_1\rightarrow 1^3P_2\gamma$&149&6.3&7.59&17.7&14.8&14.5\\
$2^1S_0\rightarrow 1P\gamma$&170&5.0&1.05&0&1.9&6.4\\
$2^1S_0\rightarrow 1P'\gamma$&124&1.9&4.40&5.2&15.9&13.1\\
$2^3P_0\rightarrow 1^3S_1\gamma$&760&0&&21.9&16.1\\
$2P\rightarrow 1^3S_1\gamma$&776&0&&22.1&15.3\\
$2P'\rightarrow 1^3S_1\gamma$&781&0&&2.1&2.5\\
$2^3P_2\rightarrow 1^3S_1\gamma$&789&0&&25.8&19.2\\
$2P\rightarrow 1^1S_0\gamma$&847&0&& &3.1\\
$2P'\rightarrow 1^1S_0\gamma$&852&0&& &20.1\\
$2^3P_0\rightarrow 2^3S_1\gamma$&197&15.0&34.0&41.2&25.5\\

$2P\rightarrow 2^3S_1\gamma$&242&31.7&45.3&54.3&32.1\\
$2P'\rightarrow 2^3S_1\gamma$&247&48.2&10.4&5.4&5.9\\
$2^3P_2\rightarrow 2^3S_1\gamma$&255&49.5&75.3&73.8&49.4\\
$2P\rightarrow 2^1S_0\gamma$&270&47.5&13.8& &8.1\\
$2P'\rightarrow 2^1S_0\gamma$&275&68.7&90.5&&58.0\\
\hline
\end{tabular*}}
\end{table*}

\begin{table*}[!h]
\caption{\label{M1}\bf M1 transition rates for the $B_c$ meson.}
{\begin{tabular*}{\textwidth}{@{\extracolsep{\fill}}lrrrrrl@{}}
\hline
Transition&\multicolumn{1}{c}{$k_{0}$}&\multicolumn{1}{c}{This work}&\multicolumn{1}{c}{Ref. \cite{FL}}&\multicolumn{1}{c}{Ref. \cite{SA95}  }&\multicolumn{1}{c}{Ref. \cite{DR03}}&\multicolumn{1}{c}{Ref.\cite{EC94}}\\
&$\Gamma(keV)$&$\Gamma(keV)$&$\Gamma(keV)$&$\Gamma(keV)$&$\Gamma(keV)$&$\Gamma(keV)$\\
\hline
$1~ ^3S_1\rightarrow 1 ^1S_0\gamma$&71&0.059&0.190&0.060&0.073&0.135\\
$2~ ^3S_1\rightarrow 2 ^1S_0\gamma$&28&0.0017&0.043&0.010&0.030&0.029\\
\hline
\end{tabular*}}
\end{table*}

\begin{table*}[!h]
\centering
\caption{\label{fbc}\bf Comparison of predictions for the pseudo scalar decay constant of the $B_c$ meson.}
\begin{tabular*}{\textwidth}{@{\extracolsep{\fill}}lrrrrrrl@{}}
\hline
Parameter&\multicolumn{1}{c}{This work}&\multicolumn{1}{c}{Ref.\cite{WS81}}&\multicolumn{1}{c}{Ref. \cite{AM80}}&\multicolumn{1}{c}{Ref.\cite{CJ77}}&\multicolumn{1}{c}{Ref.\cite{CT96}}\\
\hline
$f_{B_c}$&439.735&500&512&479&440$\pm$20\\
\hline
\end{tabular*}
\end{table*} 
\begin{table*}[!h]
\centering
\caption{\label{T1}\bf Comparison of life time of $B_c$ meson (in ps).}
\begin{tabular*}{\textwidth}{@{\extracolsep{\fill}}lrrrrrrl@{}}
\hline
This work&\multicolumn{1}{c}{Experiment\cite{PDG}}&\multicolumn{1}{c}{Ref.\cite{AA99}}&\multicolumn{1}{c}{Ref.\cite{VA95} }&\multicolumn{1}{c}{Ref.\cite{KVV}}&\multicolumn{1}{c}{Ref. \cite{GS85}}\\
\hline
0.379&0.452$\pm 0.033$&0.47&0.55$\pm 0.15$&0.50&0.75\\
\hline
\end{tabular*}
\end{table*} 

The $c\bar{b}$ states which lie below BD threshold are stable against strong decays. However, the states which are above the BD threshold undergo two body strong decays. We have calculated strong decay widths of $c\bar{b}$ states which lie above the BD threshold using the equation (\ref{strong}). The decay widths are calculated within the $^3P_0$ pair creation model. The results are presented in Table \ref{strong1}.
\begin{table*}[!h]
\centering
\caption{\label{strong1}\bf Strong decay widths of the $B_c$ meson.}
\setlength{\tabcolsep}{2pt}
\begin{tabular}{ccccccccc}
\hline
Transition&$\Gamma(MeV)$\\
\hline
$2~ ^1P_1\rightarrow B^*+D$&54.599\\
$2~ ^3P_1\rightarrow B^{*}+D$&2.145\\
$2~ ^3P_2\rightarrow B+D$&99.386\\
$2~ ^3P_2\rightarrow B^0+D^0$&108.185\\
$2~ ^3P_2\rightarrow B^*+D$&31.247\\
$1~^3D_2\to B^*+D$&0.198\\
$1~^3D_2\to B^*_s+D_s$&5.837\\
$1~^3D_2\to B^*+D^*$&2.123\\
$1~^3D_2\to B^*_s+D^*_s$&20.885\\
\hline
\end{tabular}
\end{table*}
\section{Conclusions}
\label{sec:C}
From the study of mass spectra and  decay properties of $c\bar{b}$ states using a non relativistic quark model with coupled channel effects  we draw the following conclusions
\begin{enumerate}[i)]
\item Our results for  mass spectra for $c\bar{b}$ states with coupled channel effects included for ground states agree within a few MeV, when compared to other theoretical models. For calculation of mass spectrum, the coupled channel effects are notably visible.
\item Our calculated value of the hyperfine splitting of the ground state  vector and pseudo scalar $c\bar{b}$ states 71 MeV, agree with the value predicted  by Penin et al,  $M(B^*_c)-M(B_c)=50\pm 17(th)~ \rm{MeV}$\cite{PN04}.
\item The ground state $B_c$ and  $B^*_c$ meson masses lie within the ranges  $6194~ \rm{MeV}<M_{B_c}<6292~\rm{MeV}$ and $6284~\rm{MeV}<M_{B^*_c}<6357~\rm{MeV}$ as quoted by Kwong and Rosner\cite{WJ88}. 
\item While calculating $M1$ hindered transition rates, we find relativistic effects play an important role. The zero rates of hindered transitions are due to wavefunction orthogonality. The inclusion of the relativistic effects may increase the values of hindered transition rates.
\item We find, our calculated $E1$ decay rates are in good agreement with the other theoretical model calculations. The differences found in decay rates are ascribed to differences in mass predictions, wavefunction effects and mixing angles.
\item Branching ratio for $b$-quark decays is 53\% , for $c$-quark decays 42\%  and for annihilation channel it is 5\% in estimating the weak decay widths. 
\item The life time of $c\bar{b}$ state, $f_{B_c}$ and strong decay widths predicted in this work are found to be in good agreement with experimental values as well as with other theoretical predictions.
\end{enumerate}
The NRQM in this study has proven successful in describing $B_c$ meson properties. All the observed states can be successfully accommodated in our model.
\begin{center}
\textbf{Acknowledgements}
\end{center}
One of the authors (APM) is grateful to BRNS, DAE, India for granting the project and JRF (37(3)/14/21/2014BRNS).


\bibliography{mybib}

\end{document}